\documentclass[12pt]{article}

\usepackage{graphicx}
 \usepackage{amssymb}
 \usepackage{amsmath}
 \usepackage{amsfonts}
 \usepackage{color}
 
\begin{document}

  \vspace{2cm}

  \begin{center}
    \font\titlerm=cmr10 scaled\magstep4
    \font\titlei=cmmi10 scaled\magstep4
    \font\titleis=cmmi7 scaled\magstep4
  {\bf Casimir-like corrections to the classical tensions of \\ the strings and membranes}

    \vspace{1.5cm}
    \noindent{{\large Y. Koohsarian\footnote{yo.koohsarian@stu-mail.um.ac.ir}}} \\
    {\it Department of Physics, Ferdowsi University of Mashhad \\
       P.O.Box 1436, Mashhad, IRAN} \\
     \vspace{0.3cm}
     \noindent{{ \large A. Shirzad\footnote{shirzad@ipm.ir}}}\\
    {\it Department of Physics, Isfahan University of Technology \\
       P.O.Box 84156-83111, Isfahan, IRAN,\\
       School of Physics, Institute for Research in Fundamental Sciences (IPM),\\
       P. O. Box 19395-5531, Tehran, IRAN} \\
  \end{center}
  \vskip 2em

\begin{abstract}
We find the Casimir-like energies for strings and membranes. We show that the related Casimir forces can be interpreted as quantum corrections to the classical tensions of the strings and membranes. We see that these corrections always increase the tensions of the circular string as well as spherical membrane, while for the straight string, rectangular and cylindrical membranes, these Casimir forces may increase or decrease the tensions. So we find that the quantum vacuum can break the (tensional) isotropy of the rectangular and cylindrical membranes.  Also obtaining the nonzero-temperature Casimir energy,  we find relations for the tensions at nonzero temperature.
\end{abstract}

 \textbf{Keywords}: Quantum vacuum effects, Casimir energy, strings, Membranes, Quantum-corrected tension, Nonzero-temperature Casimir force, Tensional isotropy, Tension at nonzero temperature

\section{Introduction}

It is well known that the quantum zero-point oscillations of a field, can result in some observable quantum effects such as the famous Casimir force. In fact, this force arises from the constraints on the normal modes of the vacuum state, due to some conditions imposed on the fields. So, generally, this effect depends highly, not on microscopic, but on the geometrical properties of the field system.

In this paper we consider the models of the strings and membranes. In Refs.\ \cite{CS1,CS2,CS3,CS4,CS5} (and also \cite{CS6} as a review), the Casimir effect for uniform and nonuniform piecewise strings has been studied. Also in Refs.\ \cite{CBSM1,CBSM2} the authors have found  relations for quantum vacuum induced forces between some beads on strings and flat membranes. Here we find  a direct physical interpretation of Casimir forces for strings and membranes at zero as well as nonzero temperature, that has not been given in other similar works. In section 2, we study the Casimir effect for a circular as well as straight string, and in section 3, for a rectangular, cylindrical and spherical membrane. Using the classical Lagrangian we find the classical equation of motion for the displacement field. Hence one can find a harmonic oscillatory expansion (for each transverse direction) of the fields. Taking this oscillatory modes as the quantum  oscillators, we find the vacuum state energy. At last, in section 4, we find the vacuum energy of these structures at nonzero temperature, by obtaining a mode expansion for the free energy.

\section{The quantum-corrected tension of the strings}
\subsection{Straight strings}

The classical solutions for small transverse vibrations of  straight strings are well known (the longitudinal vibrations can be neglected for very thin strings). This solutions for example for Dirichlet boundary condition at both ends (denoted as DD) are obtained as  superposition of independent classical harmonic oscillatory modes with frequencies $\upsilon_0 \frac{n\pi}{a}$ in which ``$a$'' is the length of the string and  $\upsilon_0$ is known as the velocity of the wave in the string, $\upsilon_0 = \sqrt{ \frac{\sigma_0}{\mu_0}}$ in which  the constants $\mu_0$ and $\sigma_0$ are the linear density (mass per unit length) and the classical tension of the string respectively. After quantization  we can take these classical oscillation modes to be quantum oscillators. So the vacuum energy of the string is just the summation over the zero-point energies of the quantum harmonic oscillators of two transverse directions
\begin{equation}
E_{\textrm{vac}}=2\sum_{n=1}^{\infty}\frac{1}{2} \hbar\upsilon_0 \frac{n\pi}{a} = \frac{\pi \hbar\upsilon_0}{a} \sum_{n=1}^{\infty} n. \label{eq-1}
\end{equation}
 To obtain the physical finite part of the vacuum energy, i.e.\ the Casimir energy, we regularize the above series using the known Abel-Plana formula (see for example \cite{math1}):
\begin{eqnarray}
\sum_{n=1}^{\infty} n &=& \int_0^{\infty} u du - 2 \int_0^{\infty} \frac{u du}{\exp[2\pi u]-1} \nonumber \\
&=& \int_0^{\infty} u du -\frac{1}{12}.
 \label{eq-2}
\end{eqnarray}
So using Eq.\ \eqref{eq-1} one obtains
\begin{equation}
E_{\textrm{vac}}= -\frac{\pi \hbar \upsilon_0}{12 a} + \frac{\pi \hbar\upsilon_0}{a} \int_0^{\infty} u du. \label{eq-3}
\end{equation}
Therefore  the Casimir energy and Casimir force are found as
\begin{eqnarray}
E_{\textrm{cas}}(a)&=&-\frac{\pi\hbar\upsilon_0}{12a} \nonumber \\
F_{\textrm{cas}}(a)&=&-\frac{\partial E_{\textrm{cas}}}{\partial{a}}= -\frac{\pi\hbar\upsilon_0}{12a^2}. \label{eq-4}
\end{eqnarray}
Note that the series in Eq.\ \eqref{eq-1} can  be directly regularized by applying the known Riemann zeta function (see e.g.\ \cite{math2}).  Now in order to find a physical interpretation of Casimir force for a string of length ``$a$'', suppose it is lengthened to $a + \delta a$. Then the  change  in the total ground state energy $\delta E_{\textrm{g}}^{\textrm{tot}}$, due to this lengthening,  can be written as
\begin{eqnarray}
\delta E_\textrm{g}^{\textrm{tot}} &=&\sigma_0 \delta a + \frac{\partial E_{\textrm{vac}}}{\partial a} \delta a \nonumber \\
&=&(\sigma_0 - F_{\textrm{cas}})\delta a, \label{eq-5}
\end{eqnarray}
where in the second line we have  absorbed the infinite part of the vacuum energy into the classical tension,
\begin{equation}
\left(\sigma_0 -\frac{\pi \hbar\upsilon_0}{a^2} \int_0^{\infty} u du \right) \rightarrow \sigma_0, \label{eq-6}
\end{equation}
as a definition for the physical tension of the string. This is quite similar to the renormalization of classical \emph{bare} parameters in the framework of the quantum field theory. Now  logically  we can think of $ \sigma_0 - F_{\textrm{cas}} $ as the quantum-corrected tension of the string
\begin{eqnarray}
\sigma_{\textrm{c}}(a) &\equiv & \sigma_0 - F_{\textrm{cas}}(a) \nonumber \\ &=& \sigma_0 + \frac{\pi\hbar}{12 a^2}\sqrt{\frac{\sigma_0}{\mu_0}} \hspace{1 cm} \textrm{for DD condition}. \label{eq-7}
\end{eqnarray}
So the infinite part of the vacuum energy can be absorbed as a renormalization term into the (bare) tension, while the finite part (i.e.\ the Casimir energy) gives a quantum correction to the tension of the string. Note that the vacuum energy can also contain constant terms (i.e.\ terms independent from the length parameter $a$ of the string), but these terms would not be observable and do not affect the quantum correction term, because the correction is from the Casimir force (derivative with respect to $a$) not the Casimir energy. However the uniqueness condition for the renormalized ground state energy $E_g^{\textrm{ren}}$, can be considered as
\begin{equation}
E_{g}^{\textrm{ren}} \rightarrow 0 \hspace{1 cm} \textrm{if} \hspace{1 cm} a\mu_0 \rightarrow \infty, \label{eq-8}
\end{equation}
because for a very massive ($\mu_0 a \rightarrow \infty$) string, the vacuum oscillations are expected to be negligible.  As we see from Eq.\ \eqref{eq-7}, for DD condition the string tension increases due to the quantum oscillations  of the vacuum state. The same results could be found for Neumann boundary conditions at both ends (NN) \footnote{In the case of the NN condition or the periodical condition  (for example for the circular string), the solution  has also two non-oscillation terms that correspond to translational motion of the system. In the rest frame of the system, these translational terms can simply be neglected.}. For DN boundary condition, the vacuum energy would be simply obtained as
\begin{equation}
E_{\textrm{vac}}= \frac{\pi \hbar\upsilon_0}{ a} \sum_{n=0}^{\infty} \left(n+ \frac{1}{2} \right). \label{eq-9}
\end{equation}
The above series can be regularized directly  using the  Hurwitz zeta function $\zeta_{\textrm{H}}$ (see for example \cite{math2}) as
\begin{equation}
\sum_{n=0}^{\infty} \left(n+ \frac{1}{2} \right) =\zeta_{\textrm{H}} (-1, 1/2) = \frac{1}{24}. \label{eq-10}
\end{equation}
So we obtain
\begin{eqnarray}
&& E_{\textrm{cas}} (a)= \frac{\pi \hbar \upsilon_0}{24a} \nonumber \\
&& F_{\textrm{cas}} (a)= \frac{\pi \hbar \upsilon_0}{24 a^2}. \label{eq-11}
\end{eqnarray}
Note that for DN condition, contrary to DD and NN conditions,  the Casimir force has a positive value, thus the Casimir-like correction decreases the classical tension of the string;
\begin{equation}
\sigma_{\textrm{c}}(a)=\sigma_0 - \frac{\pi\hbar}{24 a^2}\sqrt{\frac{\sigma_0}{\mu_0}} \hspace{1 cm} \textrm{for DN condition}. \label{eq-12}
\end{equation}

\subsection{Circular string}

Consider a circular string with negligible thickness. For small  circumference-vibrations of the string compared to the string radius ``$r$'', the radius can be assumed to be constant. Then the Lagrangian can be written as
\begin{eqnarray}
L=\int_0^{2\pi} d\theta \Big(\frac{\mu_0 r}{2} \left[\left(\partial_t R\right)^2  + \left(\partial_t Z\right)^2 \right]  - \frac{\sigma_0}{2r}\left[\left(\partial_{\theta} R\right)^2 + \left(\partial_{\theta}Z\right)^2 \right]\Big), \label{eq-13}
\end{eqnarray}
 in which $\theta$ is polar coordinate, and also $R(\theta,t)$ and $Z(\theta,t)$ are configuration fields equivalent to radial and transverse vibratory displacements, respectively. The equation of motion would be $[\partial_t^2-(\frac{\upsilon_0}{r}) ^2\partial _\theta^2]R(\theta,t)=0$, with $\upsilon_0 \equiv \sqrt{\frac {\sigma_0} {\mu_0}}$, and the same equation for $Z(\theta,t)$. The solutions for $Z(\theta)$ and $R(\theta)$ can be obtained (neglecting translational-motion terms) in terms of creation and annihilation operators of independent harmonic oscillatory modes. Then the vacuum energy of the string is just the expectation value of the corresponding Hamiltonian (obtained from the above Lagrangian) for the vacuum state
\begin{eqnarray}
E_{\textrm{vac}}&=&\langle 0|H|0 \rangle \nonumber \\
&=&\hbar\upsilon_0\sum_{n=1}^{\infty}\frac{n}{r}, \label{eq-14}
\end{eqnarray}
that is just the summation over the zero-point energies of the quantum harmonic oscillatory modes. Therefore $E_{cas}(r)$ and $F_{cas}(r)$ can be found as
\begin{eqnarray}
E_{\textrm{cas}}(r)&=&-\frac{\hbar\upsilon_0}{12 r} \nonumber \\
F_{\textrm{cas}}(r)&=&-\frac{\partial E_{\textrm{cas}}}{\partial r}=-\frac{\hbar\upsilon_0}{12 r^2}. \label{eq-15}
\end{eqnarray}
Note that this time, the Casimir force is  central.  The variation $\delta E_\textrm{g}^{\textrm{tot}}$ caused by the change $2\pi \delta r$ in circumference of the string, can be written as
\begin{eqnarray}
 \delta E_\textrm{g}^{\textrm{tot}} &=&2\pi \sigma_0 \delta r +\frac{\partial E_{\textrm{cas}}}{\partial r} \delta r \nonumber \\
&=& 2\pi \left(\sigma_0+\frac{\hbar\upsilon_0}{24\pi r^2}\right)\delta r.
\label{eq-16}
\end{eqnarray}
So
\begin{equation}
\sigma_{\textrm{c}}(r)\equiv\sigma_0+\frac{\hbar}{24\pi r^2}\sqrt{\frac{\sigma_0} {\mu_0}} \label{eq-17}
\end{equation}
can be interpreted as the quantum corrected tension. As we see quantum vacuum correction always increases  the classical tension of the circular string.

\section {The quantum-corrected tension of the membranes}
\subsection{Rectangular flat membrane}

A rectangular flat membrane (Figure 1) with negligible thickness and for small surface vibrations, can be described by the known Lagrangian
\begin{equation}
L= \frac{1}{2}\int_0^b dx\int_0^a dy \left[\rho_0\left(\partial_t Z\right)^2-\tau_0\left(\partial_x Z\right)^2-\tau_0\left(\partial_y Z\right)^2\right], \label{eq-18}
\end{equation}
where the displacement field $Z(x,y,t)$ corresponds to surface vibrations and $\rho_0$ and $\tau_0$ are surface-density (mass per unit area) and  surface-tension (tension per unit length) of the membrane respectively.
\begin{figure}[!b]
\centering
\includegraphics [width=5 cm, height= 7  cm]{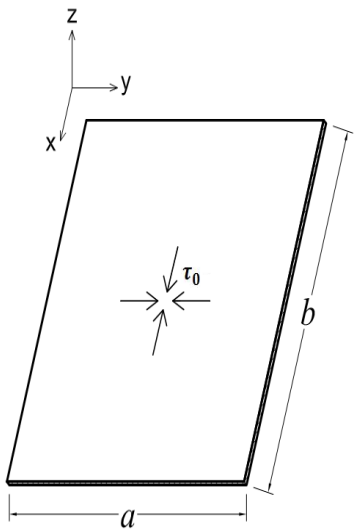}
\caption{Rectangular flat membrane}
\end{figure}
The equation of motion is obtained as the wave equation $(\partial_t^2-\upsilon_0^2\partial_x^2-\upsilon_0^2 \partial_y^2) Z(x,y,t)=0$, in which $\upsilon_0=\sqrt{\frac{\tau_0}{\rho_0}}$. For DD  boundary condition, the solution can be written (up to coefficients) as
\begin{eqnarray}
Z_{n,m} (x,y,t)\sim   \sin (\frac{n \pi}{b}x) \sin (\frac{m \pi}{a}y) \exp(-i \alpha_{n,m} t) + \textrm{c.c.} \label{eq-19}
\end{eqnarray}
in which $\alpha_{n,m} \equiv \upsilon_0\sqrt{(\frac{n\pi}{b})^2 +(\frac{m\pi}{a})^2}$ and the ``c.c.'' stands for the complex conjugate terms. So the vacuum energy would be obtained as
\begin{equation}
E_{\textrm{vac}}=\frac{\hbar\upsilon_0}{2}\sum_{n,m=1}^\infty \sqrt{(\frac{n\pi}{b})^2+(\frac{m\pi}{a})^2}. \label{eq-20}
\end{equation}
This series can be regularized applying twice the  Abel-Plana formula, and the infinite terms can be absorbed as some renormalization terms into the bare tension of the membrane. Then the Casimir energy, and the Casimir forces of $x$- and $y$-direction are found as ($b\geq a$)
\begin{eqnarray}
&&E_{\textrm{cas}}(a,b)= \hbar \upsilon_0 \left(\frac{\pi}{48 a} -\frac{b}{16\pi a^2} \zeta_R(3)-\frac{1}{a} G(\frac{b}{a})\right)  \nonumber \\
&&F^{\textrm{cas}}_x(a,b)=-\frac{\partial E_{\textrm{cas}}}{\partial b} = \hbar\upsilon_0 \left(\frac{ \zeta_R(3)}{16\pi}+ G\,'(\frac{b}{a}) \right)\frac{1}{a^2}
 \nonumber \\
&& F^{\textrm{cas}}_y(a,b)=-\frac{\partial E_{\textrm{cas}}}{\partial a} = \hbar\upsilon_0 \Big[\frac{\pi}{48 a^2} - \frac{b}{8\pi a^3}\zeta_R(3) -\frac{1}{a^2} G(\frac{b}{a}) - \frac{b}{a^3} G\,'(\frac{b}{a}) \Big]
\label{eq-21}
\end{eqnarray}
where $\zeta_\textrm{R}$ denotes the Riemann zeta function, $G(\frac{b}{a})\equiv\frac{1}{2} \sum_{n,m=1}^\infty \frac{n}{m} K_1(2\pi n m\frac{b}{a})$ in which $K$ is Bessel function of the second kind, and the ``$prime$'' denotes the derivative with respect to the argument; $G\,'(b/a)\equiv \frac{\partial G(b/a)}{\partial(b/a)}$. With numerical computations, one can find $0 < G(b/a)\leq G(1)\approx 0.000495$ and $-0.00339 \approx G\,'(1)\leq G\,'(b/a)< 0$, thus $F^{\textrm{cas}}_x$ is always positive (i.e.\ repulsive), but $F^{\textrm{cas}}_y$ may be  negative (i.e.\ attractive). In practice, the terms $G(b/a)$ and $G\,'(b/a)$ can be ignored with a good degree of accuracy, therefore we find that $F^{\textrm{cas}}_y$ is repulsive if $1\leq \frac{b}{a}\lesssim 1.37$ and is attractive if $\frac{b}{a}\gtrsim 1.37$. Note that our relations in Eq.\ \eqref{eq-21} are in agreement with comparable expressions of other parallel works (see for example section 4 of \cite{NDCE}), however, our numerical results for the Casimir forces is not given in other works. Now in order to interpret these results more physically, we write
\begin{eqnarray}
\delta E_\textrm{g}^{\textrm{tot}} &=&(b\, \tau_0)\delta a+ \frac{\partial E_{\textrm{cas}}}{\partial a}\delta a + (a\, \tau_0)\delta b+ \frac{\partial E_{\textrm{cas}}}{\partial b}\delta b \nonumber \\ &=& b\big(\tau_0 - \frac{1}{b} F^{\textrm{cas}}_y \big)\delta a + a\big(\tau_0 - \frac{1}{a} F^{\textrm{cas}}_x\big)\delta b, \label{eq-22}
\end{eqnarray}
where \ $a\tau_0$ \ and \ $b\tau_0$\ are total tensions of $x$- and $y$-directions respectively.  Then we can take the corrected tensions as
\begin{eqnarray}
\tau_x(a,b)&\equiv & \tau_0-\frac{1}{a}F^{cas}_x(a,b) \nonumber \\ &=& \tau_0 -\hbar\sqrt{\frac{\tau_0}{\rho_0}} \left(\frac{ \zeta_R(3)}{16\pi}+ G\,'(\frac{b}{a}) \right)\frac{1}{b a^2}
, \nonumber \\ \tau_y (a,b)&\equiv& \tau_0-\frac{1}{b}F^{cas}_y(a,b)\nonumber \\ &=&\tau_0 -\hbar\sqrt{\frac{\tau_0}{\rho_0}} \Big[\frac{\pi}{48 b a^2} - \frac{\zeta_R(3)}{8\pi a^3} - \frac{1}{b a^2} G(\frac{b}{a})-\frac{1}{a^3} G\,'(\frac{b}{a})\Big]
 \label{eq-23}
\end{eqnarray}
So $\tau_x$  always decreases, i.e.\ tensional stableness of the membrane at $x$-direction decreases while, $\tau_y$ decreases  if $1\leq \frac{b}{a} \lesssim 1.37$, and  increases  when $\frac{b}{a}\gtrsim 1.37$. As is seen $\tau_x\neq \tau_y$ that is, the vacuum oscillations generally  break the (tensional) isotropy of a rectangular membrane.  For a square membrane $(b=a)$, we find
\begin{eqnarray}
 F^{cas}_x(a)&=& F^{cas}_y(a)\approx  0.02 \  \frac{\hbar\upsilon_0}{a^2} \nonumber \\ \tau_x(a)
 &=& \tau_y(a)\ \approx  \tau_0 - 0.02 \ \hbar\sqrt{\frac{\tau_0}{\rho_0}} \frac{1}{a^3}, \label{eq-24}
\end{eqnarray}
so quantum corrections always decrease the  surface tension of the square membrane. We see that the vacuum oscillations respect the isotropy of a square membrane. For NN boundary condition, one can write
\begin{eqnarray}
Z_{n,m} (x,y,t)\sim   \cos (\frac{n \pi}{b}x) \cos (\frac{m \pi}{a}y) \exp(-i \alpha_{n,m} t)+ \cos (\frac{n \pi}{b}x) \exp(-i \upsilon_0 \frac{n \pi}{b} t) \nonumber \\ +  \cos (\frac{m \pi}{a}y) \exp(-i \upsilon_0 \frac{m \pi}{a} t) +  \textrm{c.c.} \label{eq-25}
\end{eqnarray}
in which $\alpha_{n,m}$ is defined as before. So the vacuum energy is obtained as
\begin{eqnarray}
E_{\textrm{vac}}=\frac{\hbar\upsilon_0}{6}\Big(\sum_{n,m=1}^\infty \sqrt{(\frac{n\pi}{b})^2+(\frac{m\pi}{a})^2}  + \sum_{n=1}^{\infty} \frac{n\pi}{b} +\sum_{m=1}^{\infty} \frac{m\pi}{a} \Big), \label{eq-26}
\end{eqnarray}
and after regularization we find the Casimir energy:
\begin{eqnarray}
E_{\textrm{cas}}(a,b)= -\frac{\hbar \upsilon_0}{3} \Big[\frac{\pi}{48} \frac{1}{a}+\frac{\pi}{24 b} + \frac{b}{16\pi a^2} \zeta_R(3) +\frac{1}{a}G(\frac{b}{a})\Big]. \label{eq-27}
\end{eqnarray}
As is seen the Casimir energy of the NN condition is different from that of DD condition.

\subsection{Cylindrical membrane}

 The Lagrangian for small surface-vibrations of a cylindrical membrane (see figure 2) can be written as
\begin{eqnarray}
L= \frac{1}{2}\int_0^{2\pi} d\phi \int_0^l dz \big[r \rho_0 \left(\partial_t R\right)^2 - r \tau_0  \left(\partial_z R\right)^2 -\frac {\tau_0}{r} \left(\partial_{\phi} R\right)^2\big],
\label{eq-28}
\end{eqnarray}
 where the field $R(z,\phi,t)$ corresponds to vibrational displacement (here for small vibrations, ``$r$'' is taken to be constant).
 \begin{figure}[!b]
\centering
\includegraphics [ width=8 cm, height= 5  cm]{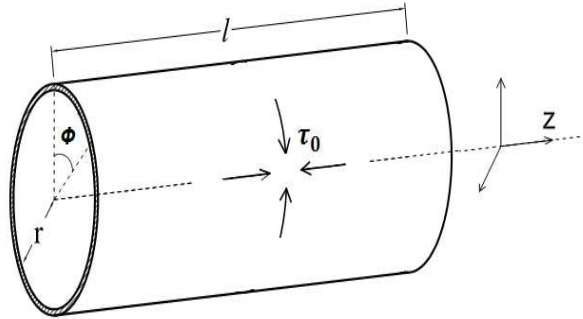}
\caption{Cylindrical membrane}
\end{figure}
Now we have a boundary condition for longitudinal variable ``$z$'', and a periodical condition for azimuthal variable $\phi$. For DD boundary condition one obtains an oscillatory mode expansion as
\begin{eqnarray}
R_{n,m} (z,\phi,t)\sim   \sin (\frac{n \pi}{ l}z) \exp(+i m\phi) \exp(-i \omega_{n,m} t)  \nonumber \\   +\sin (\frac{n \pi}{ l}z) \exp(-i m \phi) \exp(-i \omega_{n,m} t) \nonumber \\ +\sin (\frac{n \pi}{ l}z) \exp (-i \upsilon_0 \frac{n \pi}{ l} t) + \textrm{c.c.} \label{eq-29}
\end{eqnarray}
 in which $\omega_{n,m} \equiv \upsilon_0\sqrt{ (n\pi / l)^2+ (m/r)^2}$ with $\upsilon_0\equiv\sqrt{\tau_0 /\rho_0}$. Therefore the vacuum energy of the membrane would be obtained as
\begin{equation}
E_{\textrm{vac}}=\frac{\hbar\upsilon_0}{4} \Big(\sum_{n,m=1} ^\infty  \sqrt{(\frac{n\pi}{ l})^2+(\frac{m}{r})^2} + \sum_{n=1}^{\infty}\frac{n\pi}{ l} \Big)   \label{eq-30}
\end{equation}
Similarly as for rectangular membrane, we find the Casimir energy, Casimir radial and longitudinal forces as ($ l \geq \pi r$)
\begin{eqnarray}
E_{\textrm{cas}}(r, l)&=&  \frac{\hbar \upsilon_0}{2} \Big( \frac{1}{48 r} -\frac{\pi}{24  l} - \frac{ l}{16\pi^3 r^2} \zeta_R(3) -\frac{1}{\pi r} G(\frac{ l}{\pi r})\Big) \nonumber \\
F^{\textrm{cas}}_z(r, l)&=& \frac{\hbar \upsilon_0}{2}\Big(-\frac{\pi}{24 l^2} +\frac{\zeta_\textrm{R}(3)}{16\pi^3r^2} +\frac{1}{\pi^2r^2}G\,'(\frac{ l}{\pi r}) \Big)\nonumber \\
F^{\textrm{cas}}_r(r, l)&=& \frac{\hbar \upsilon_0}{2}\Big( \frac{1}{48 r^2}- \frac{ l}{8\pi^3 r^3} \zeta_\textrm{R}(3) - \frac{1}{\pi r^2}G(\frac{ l}{\pi r}) - \frac{ l}{\pi^2 r^3} G\,'(\frac{ l}{\pi r}) \Big)
\label{eq-31}
\end{eqnarray}
Then we write
\begin{eqnarray}
 \delta E_\textrm{g}^{\textrm{tot}}&=&( l\, \tau_0)(2\pi\delta r) + \frac{\partial E_{\textrm{cas}}}{\partial r}\delta r + (2\pi r\, \tau_0)\delta  l+ \frac{\partial E_{\textrm{cas}}}{\partial  l}\delta  l \nonumber \\
&=& 2\pi l\big(\tau_0 - \frac{1}{2\pi l} F^{\textrm{cas}}_r \big)\delta r + 2\pi r\big(\tau_0 - \frac{1}{2\pi r} F^{\textrm{cas}}_z \big)\delta  l, \label{eq-32}
\end{eqnarray}
 with \ $2\pi r \tau_0$\ and\ $ l \tau_0$\ as the total longitudinal and azimuthal tensions respectively. Hence the quantum-corrected tensions can be taken as
\begin{eqnarray}
\tau_{\textrm{azim}}(r, l) &\equiv & \tau_0 - \frac{1}{2\pi l}F^{\textrm{cas}}_r  (r, l)  \nonumber \\
&=& \tau_0 - \hbar \sqrt {\frac{\tau_0}{\rho_0}} \Big( \frac{1}{192 \pi l r^2}- \frac{ \zeta_R(3)}{32\pi^4 r^3}  - \frac{1}{4\pi^2  l r^2}G(\frac{ l}{\pi r}) -  \frac{1}{4\pi^3 r^3} G\,'(\frac{ l}{\pi r}) \Big) \nonumber \\
\tau_{\textrm{long}}(r, l) &\equiv & \tau_0 - \frac{1}{2\pi r}F^{\textrm{cas}}_z (r, l)
\nonumber \\ &=& \tau_0 - \hbar\sqrt{\frac {\tau_0} {\rho_0}} \Big(-\frac{1}{96 r  l^2}+\frac{\zeta_\textrm{R}(3)}{64\pi^4r^3} +\frac{1}{4\pi^3r^3}G\,'(\frac{ l}{\pi r}) \Big) \label{eq-33}
\end{eqnarray}
So $\tau_{\textrm{azim}}$ decreases if $1\leq \frac{ l}{\pi r}\lesssim 1.37$, and increases if $\frac{ l}{\pi r}\gtrsim 1.37$, while $\tau_{\textrm{long}}$  increases if $1\leq \frac{ l}{\pi r}\lesssim 2.34$ and decreases if $\frac{ l}{\pi r}\gtrsim 2.34$. As is seen the quantum corrections can also break the isotropy of the cylindrical membrane. For NN boundary condition, one can write
\begin{eqnarray}
R_{n,m} (z,\phi,t)\sim   \exp(-i m \phi) \exp\left(-i\upsilon_0 \frac{m}{r}t\right) + \exp\left(+i m\phi\right) \exp\left(-i\upsilon_0 \frac{m}{r}t\right) + \nonumber \\ \cos (\frac{n \pi}{ l}z) \exp\left(+i m\phi\right) \exp\left(-i \omega_{n,m} t\right)   + \cos (\frac{n \pi}{ l}z) \exp(-i m \phi) \exp\left(-i \omega_{n,m} t\right)
\nonumber \\  + \cos(\frac{n \pi}{ l}z) \exp\left(-i\upsilon_0 \frac{n \pi}{ l}t\right) + \textrm{c.c.} \label{eq-34}
\end{eqnarray}
 so the vacuum energy can be obtained as
\begin{eqnarray}
E_{\textrm{vac}}=\frac{\hbar\upsilon_0}{6} \Big(\sum_{n,m=1} ^\infty  \sqrt{(\frac{n\pi}{ l})^2+(\frac{m}{r})^2} + \sum_{n=1}^{\infty}\frac{n\pi}{ l} + \sum_{m=1}^{\infty}\frac{m}{r} \Big), \label{eq-35}
\end{eqnarray}
and so the Casimir energy would be found as
\begin{eqnarray}
E_{\textrm{cas}}(r, l)&=& - \frac{\hbar \upsilon_0}{3}\Big( \frac{1}{48 r}+ \frac{\pi}{24  l} + \frac{ l}{16\pi^3 r^2} \zeta_\textrm{R}(3) +\frac{1}{\pi r} G(\frac{ l}{\pi r})\Big) \label{eq-36}
\end{eqnarray}

\subsection{Spherical membrane}

The Lagrangian of a spherical thin membrane of radius ``$r$''  with small surface vibrations can be written as
\begin{eqnarray}
L= \frac{1}{2}\int_0^{2\pi} d\phi \int_0^\pi \sin\theta \, d\theta \big[ \rho_0 r^2 \left(\partial_t R\right)^2 - \tau_0 \left(\partial_\theta R\right)^2  -\frac{\tau_0}{\sin^2 \theta} \left(\partial_\phi R\right)^2\big],
\label{eq-37}
\end{eqnarray}
The equation of motion is found as
\begin{equation}
\left[\left(\frac{r}{\upsilon_0}\right)^2 \partial^2_t -\frac{1}{\sin \theta}\partial_{\theta}\left(\sin \theta \partial_{\theta}\right)- \frac{1}{\sin^2\theta} \partial_{\phi}^2\right]R(\theta,\phi,t)=0. \label{eq-38}
\end{equation}
The angular part of this equation leads just to the generalized Legendre equation, therefore the  solution is written in terms of spherical harmonics
 \begin{eqnarray}
R_{k,m}(\theta,\phi,t)\sim  \exp{\left(-i\frac{\upsilon_0}{r} t \sqrt{k(k+1)}\right)} \mathcal{Y}_{k,m}(\theta,\phi)+ \textrm{c.c.} \label{eq-39}
\end{eqnarray}
in which, $\mathcal{Y}_{k,m}$ are the spherical harmonics. Then after some calculations one can find the vacuum energy as
\begin{equation}
E_{\textrm{vac}}= \frac{\hbar\upsilon_0}{2r} \sum_{k=1}^{\infty}(2k+1)\sqrt{k(k+1)}. \label{eq-40}
 \end{equation}
The above series can be regularized using the known Euler-Maclaurin summation formula, see e.g.\ \cite{math3}. Here, ignoring irrelevant infinite  terms, this formula can be approximated as
\begin{eqnarray}
&&\sum_{k=1}^{\infty} f(k) \simeq  \int_1^{\infty} f(k) d k + \frac{1}{2} f(k=1) -\frac{1}{12} \left(\frac{\partial f}{\partial k}\right)_{k=1} + \frac{1}{720} \left(\frac{\partial^3 f}{\partial k^3}\right)_{k=1} \label{eq-41}
\end{eqnarray}
where $f$ is an analytic function in integration region, and in the right-hand side $k$ is taken as a continuous variable. So the series in the vacuum energy \eqref{eq-40} can be regularized as
\begin{eqnarray}
\sum_{k=1}^{\infty}\left(2k+1 \right)\sqrt{k(k+1)} \simeq -\int_0^1 d k \left(2k+1 \right)\sqrt{k(k+1)} + \frac{5867}{2560 \sqrt{2}} \approx -0.265 \label{eq-42}
\end{eqnarray}
The above result is in agreement with corresponding  result in Ref.\  \cite{FL} obtained from  zeta function regularization. Then we find the Casimir energy and Casimir force:
\begin{eqnarray}
E_{\textrm{cas}}(r)&\approx & - 0.133\frac{\hbar \upsilon_0}{r} \nonumber \\ F_{\textrm{cas}}(r)&=& -\frac{\partial E_{\textrm{cas}}}{\partial r}\approx - 0.133 \frac{\hbar \upsilon_0}{r^2},
 \label{eq-43}
\end{eqnarray}
 As we see the Casimir Force for a spherical membrane is always central. Again it is desirable to find an appropriate physical explanation for this force. We write
\begin{eqnarray}
\delta E_\textrm{g}^{\textrm{tot}} &=& \tau_0 \left(2r\delta r\int_0^{2\pi}d\phi \int_{-1}^{1}d(\cos\theta)\right) +\frac{\partial E_{\textrm{cas}}}{\partial r} \delta r  \nonumber \\
&=& 8\pi r\left(\tau_0 - \frac{1}{8\pi r} F_{\textrm{cas}}(r)\right)\delta r.  \label{eq-44}
\end{eqnarray}
This relation can be understood someway noting that the change in the spherical area of the membrane, due to the infinitesimal change in radius $\delta r$, equals $\delta(4 \pi r^2)= 8\pi r \delta r$. Now the quantum corrected (surface) tension can be taken as
\begin{equation}
\tau(r)\equiv \tau_0 - \frac{1}{8\pi r} F_{cas}(r) \approx\tau_0+ 0.00527 \, \hbar\sqrt{\frac{\tau_0} {\rho_0}}\frac{1}{r^3},  \label{eq-45}
\end{equation}
so the Casimir-like correction always increases the surface tension of spherical membrane. Also we see that the vacuum oscillations do not break the isotropy of the classical tension of spherical membrane.

\section{The tension at nonzero temperature}

So far we have not considered the thermal effects in obtaining the Casimir forces, while a physical system actually involves thermal excitations. In the framework of the statistical mechanic, as we know, a thermal equilibrated system is considered  as an ensemble of states having the same temperature,   with a probability distribution. The regularized free energy of this ensemble then provides the Casimir energy at nonzero temperature \cite{ACE}. Here we analyze the  thermal equilibrated systems applying the known Matsubara formalism. As we know, the thermodynamic of a system is described by a partition function having the functional integral representation $Z\sim\int D\Psi \exp(-S_\textrm{E}[\Psi])$ in which, $\Psi$ is the configuration space field and $S_\textrm{E}[\Psi]$ is the Euclidean action of the system (reminding that $S_\textrm{E}[\Psi]$ is obtained by Wick rotation of time coordinate: $t\rightarrow-it$).

\subsection{strings at nonzero temperature}

The Euclidean action of the straight string can be written as
 \begin{eqnarray}
 S_E[Y]= i\int dt \int dx \left[Y(x,t)\Delta Y(x,t)\right] \ \ \ \textrm{with} \hspace{0.5 cm} \Delta \equiv -\frac{1}{2}\mu_0 \partial_t^2 - \frac{1}{2}\sigma_0 \partial_x^2 \label{eq-46}
 \end{eqnarray}
So the partition function will be
\begin{equation}
\mathcal{Z}=\int DY \exp(iS_E)\sim\left(\det \Delta \right)^{-1}. \label{eq-47}
\end{equation}
 Then ignoring a constant term, the free energy is obtained as \begin{equation}
\mathcal{F}\equiv -k_B T \ln{Z} = k_B T \, \mathrm{Tr}(\ln{\Delta}), \label{eq-48}
 \end{equation}
in which $k_B$ is  the Boltzmann constant and ``$T$'' is the temperature of the thermal equilibrated system (here the string). Utilizing the Matsubara formalism one can write the field modes (for Dirichlet condition) as
 \begin{equation}
Y_{nj}(x,t)= \frac{1}{\sqrt{2\pi}}\exp(-i\xi_j t) \sin\left(\frac{n\pi}{a} x \right), \label{eq-49}
 \end{equation}
 which contains the Matsubara frequencies $\xi_j\equiv 2\pi j k_B T /\hbar$. Since
 \begin{equation}
 \Delta Y_{nj}=\left[\frac{1}{2}\mu_0 \xi_j^2 +\frac{1}{2}\sigma_0 \left(\frac{n\pi}{a}\right)^2 \right]Y_{nj}, \label{eq-50}
 \end{equation}
dropping an infinite constant term, the free energy of the string is found as
 \begin{equation}
\mathcal{F}(a,T)=k_B T\sum_{j=-\infty}^{\infty} \sum_{n=1} ^{\infty} \ln\left[\xi_j^2 + \frac{\sigma_0}{\mu_0}\left(\frac{n\pi}{a}\right)^2 \right]. \label{eq-51}
\end{equation}
Such an expression can be simplified as the sum of a zero-temperature part $\mathcal{F}_{0}$ and a thermal correction term $\mathcal{F}_T$  (see for example chapter 5 of Ref.\ \cite{ACE}) as
\begin{eqnarray}
&&\mathcal{F}(a,T) =\mathcal{F}_{0} + \mathcal{F}_T \nonumber \\
&&\mathcal{F}_{0}=\hbar\upsilon_0\sum_{n=1} ^{\infty}\frac{n \pi}{a}\nonumber \\
&&\mathcal{F}_T = 2k_B T \sum_{n=1}^{\infty} \ln\left[1-\exp\left(-\frac{\hbar \upsilon_0} {k_B T}\frac{n\pi}{a}\right)\right].
\label{eq-52}
\end{eqnarray}

The above calculation may be explained in a simple physical way: As we realized before, the strings (as well as membranes) can be considered as a superposition of independent harmonic oscillatory modes. So a thermal equilibrated string of temperature $T$ can be regarded as a superposition of infinite number of independent quantum harmonic oscillators, each of which being equilibrated at the temperature $T$. For a harmonic oscillator of temperature $T$, the partition function is given as
  \begin{equation}
\Lambda_n= \exp\left(-\frac{\hbar}{2 k_B T} \omega_n \right)\left[1- \exp \left(-\frac{\hbar}{k_B T} \omega_n \right) \right]^{-1}, \label{eq-53}
\end{equation}
 in which $\omega_n$ is the oscillation frequency. Taking $\omega_n= \upsilon_0 \frac{n \pi}{a}$ we can assign a free energy for each transverse direction of each mode of the straight string as \begin{eqnarray}
\mathcal{F}_n= -k_B T \ln \Lambda_n =\frac{\hbar \upsilon_0 }{2}\frac{n\pi}{a}  + k_B T \ln \left[1-\exp\left(-\frac{\hbar \upsilon_0}{k_B T} \frac{n\pi}{a}\right)\right]. \label{eq-54}
\end{eqnarray}
 The free energy of the string is then obtained just by summing over the free energies of the mode spectrum of all transverse directions  $\mathcal{F}=2\sum_{n=1}^{\infty}\mathcal{F}_n$, and so Eqs.\ \eqref{eq-52} will be directly obtained.

The zero-temperature part $\mathcal{F}_0$ gives (after regularization) just the Casimir energy of the string:
\begin{equation}
\mathcal{F}_0= -\frac{\pi \hbar \upsilon_0}{12 a}=E_{\textrm{cas}}(a). \label{eq-55}
\end{equation}
 As a result, in the framework of the Casimir effect the  free energy $\mathcal{F}$ is taken as the nonzero-temperature Casimir energy \cite{NDCE,ACE}.  So the nonzero-temperature Casimir energy of the string can be written as
 \begin{eqnarray}
\mathcal{F}_{\textrm{cas}}(a,T) \equiv \mathcal{F}(a,T)
 = E_{\textrm{cas}}(a) + 2k_B T \sum_{n=1}^{\infty} \ln \left[1-\exp\left(-\frac{\hbar \upsilon_0}{k_B T} \frac{n\pi}{a}\right)\right],  \label{eq-56}
\end{eqnarray}
and so the nonzero-temperature Casimir force can be given as
\begin{equation}
 F_{\textrm{cas}}(a,T) \equiv -\frac{\partial \mathcal{F}_{\textrm{cas}}}{\partial a} \label{eq-57}
\end{equation}
Physically we expect that for large $a$ (i.e.\ for a long string) the nonzero-temperature Casimir force tends to zero. But using Abel-Plana formula one can find for large $a$
 \begin{equation}
 \sum_{n=1}^{\infty} \ln \left[1-\exp\left(-\frac{\hbar \upsilon_0}{k_B T} \frac{n\pi}{a}\right)\right]= -\frac{\pi k_B T}{6 \hbar \upsilon_0} a + O\left[\ln a\right]. \label{eq-58}
 \end{equation}
From the first term in the right-hand side of the above relation, $F_{\textrm{cas}}(a,T)$ would contain a nonzero term  at the limit  $a \rightarrow \infty$.  So we should subtract this term, to obtain the physical Casimir energy at nonzero temperature:
 \begin{eqnarray}
\mathcal{F}^{\textrm{cas}}(a,T) = -\frac{\pi \hbar \upsilon_0}{12 a} + 2k_B T \sum_{n=1}^{\infty} \ln \left[1-\exp\left(-\frac{\hbar \upsilon_0}{k_B T} \frac{n\pi}{a}\right)\right] +\frac{\pi a}{3 \hbar \upsilon_0} (k_B T)^2 \label{eq-59}
\end{eqnarray}
Now the nonzero-temperature Casimir force can be found as
\begin{eqnarray}
F_{\textrm{cas}}(a,T) \equiv -\frac{\partial \mathcal{F}_{\textrm{cas}}}{\partial a} = F_{\textrm{cas}}(a)\left(1- \sum_{n=1}^{\infty} \frac{24 n\exp\left( -n\theta/T \right)}{1- \exp\left( -n\theta/T \right)}+ 4 \pi^2 \left(\theta/T\right)^{-2}\right),
 \label{eq-60}
\end{eqnarray}
in which $ \theta \equiv \pi \hbar \upsilon_0 /a k_B$, and we have used Eq.\ \eqref{eq-4}. Therefore following  Eq.\ \eqref{eq-7}, one can find the corrected tension of the string at nonzero temperature:
\begin{equation}
\sigma(a,T) \equiv \sigma_0 + \frac{\partial \mathcal{F}^{\textrm{cas}}}{\partial a}= \sigma_0 + \frac{ \pi \hbar \upsilon_0} {12 a^2}\left(1- \sum_{n=1}^{\infty} \frac{24 n\exp\left( -n\theta/T \right)}{1- \exp\left( -n\theta/T \right)}+ 4 \pi^2 \left(\theta/T\right)^{-2}\right). \label{eq-61}
\end{equation}
At low temperature we can approximate the series in  Eq.\ \eqref{eq-61} with its first term, so we can write
\begin{equation}
\sigma(a,T) \approx \sigma_0 + \frac{ \pi \hbar \upsilon_0} {12 a^2}\left(1-24\exp\left( -\theta/T \right) + 4 \pi^2 \left(\theta/T\right)^{-2}\right); \hspace{1 cm} T\ll \theta. \label{eq-62}
\end{equation}
The corresponding results for the circular string  can be found in a quite similar way.

\subsection{Membranes at nonzero temperature}

 The temperature corrections for membranes are attained similarly as we did for strings. For a rectangular flat membrane (with Dirichlet boundary condition) we have
\begin{eqnarray}
&&\mathcal{F}_T= k_B T\sum_{n=1}^{\infty}\sum_{m=1}^{\infty} \ln\left(1- \exp\left[-\frac{\hbar} {k_B T} \alpha_{n,m} \right]\right) \nonumber \\
&&\alpha_{n,m} \equiv \upsilon_0 \sqrt{\left(\frac{n \pi}{a}\right)^2+ \left(\frac{m \pi}{b}\right)^2} \label{eq-63}
\end{eqnarray}
 so
\begin{eqnarray}
&& \hspace{3 cm} \mathcal{F}_{\textrm{cas}}(a,b,T)= E_{\textrm{cas}}(a,b) + \mathcal{F}_T \nonumber \\ && F_x^{\textrm{cas}}(a,b,T)= -\frac{\partial \mathcal{F}_{\textrm{cas}}}{\partial b} \hspace{1 cm} \textrm{and} \hspace{1 cm} F_y^{\textrm{cas}}(a,b,T)= -\frac{\partial \mathcal{F}_{\textrm{cas}}}{\partial a} \label{eq-64}
\end{eqnarray}
with $E_{\textrm{cas}}(a,b)$ given by Eq.\ \eqref{eq-21}. Then applying twice the Abel-Plana formula, we find for large values of $a$ and $b$
\begin{eqnarray}
\sum_{n,m=1}^{\infty} \ln\left(1- \exp\left[-\frac{\hbar} {k_B T} \alpha_{n,m}\right]\right) \approx \frac{\pi}{12} \frac{k_B T}{\hbar\upsilon_0} \left(a+b\right)
 -\frac{\zeta_R(3)}{2 \pi} \left(\frac{k_B T}{\hbar\upsilon_0}\right)^2 a b . \label{eq-65}
 \end{eqnarray}
Thus requiring  $F_x^{\textrm{cas}}$ and $F_y^{\textrm{cas}}$ to tend to zero at large $b$ and $a$ respectively, we will find the physical $\mathcal{F}_{\textrm{cas}}$ as
\begin{eqnarray}
\mathcal{F}_{\textrm{cas}}(a,b,T)= E_{\textrm{cas}}(a,b) + k_B T\sum_{n,m=1}^{\infty} \ln\left(1- \exp\left[-\frac{\hbar} {k_B T} \alpha_{n,m}\right]\right) \nonumber \\
-\frac{\pi}{12} \frac{\left(k_B T\right)^2}{\hbar\upsilon_0} \left(a+b\right)+ \frac{\zeta_R(3)}{2\pi}\frac{(k_B T)^3}{(\hbar \upsilon_0)^2}a b. \label{eq-66}
\end{eqnarray}
Then the corrected tensions would be given as
\begin{eqnarray}
\tau_x(a,b,T)= \tau_0 -\frac{1}{a} F_x^{\textrm{cas}}(a,b,T) = \tau_0 + \frac{1}{a} \frac{\partial\mathcal{F}_{\textrm{cas}}}{\partial b} \nonumber \\ \tau_y(a,b,T)= \tau_0 -\frac{1}{b} F_y^{\textrm{cas}}(a,b,T) = \tau_0 + \frac{1}{b} \frac{\partial\mathcal{F}_{\textrm{cas}}}{\partial a}
\label{eq-67}
\end{eqnarray}
Keeping the first term of the series in Eq.\ \eqref{eq-66}, we find a low-temperature correction to the tension of for example $x$-direction as
\begin{eqnarray}
-\frac{ \pi \hbar \upsilon_0}{ a b^2}
\frac{\exp\left[-(\eta/T) \sqrt{1+(b/a)^2}\right]}{\sqrt{1+(b/a)^2}}- \frac{\pi}{12 a} \frac{ (k_B T)^2}{\hbar \upsilon_0}+ \frac{\zeta_R(3)}{2\pi}\frac{(k_B T)^3}{(\hbar \upsilon_0)^2}; \hspace{1cm} T\ll \eta.
\label{eq-68}
\end{eqnarray}
in which $\eta \equiv \pi \hbar \upsilon_0/b k_B$.  Corresponding results are similar for the cylindrical membrane.

\section*{acknowledgments}
 We thank Ali Naji for reading our paper and his valuable discussions,  Emilio Elizalde for his useful comments, also Gholamreza Yazarloo and Hamid Razaghian for their helps during the numerical calculations, and Amin Khazaee for preparing figures.



\end{document}